\documentclass[aps,prb,twocolumn]{revtex4}  
\usepackage{epsfig}
\usepackage{amsmath}
\usepackage{ulem}
\newcommand{\DM}{Dzyaloshinskii-Moriya }

\begin{document}
\title{Magnon  Dispersion and Anisotropies in SrCu$_2$(BO$_3$)$_2$}
\author{Y. F. Cheng$^1$, O. C\'epas$^2$, P. W. Leung$^1$, and T. Ziman$^{3,*}$}
\affiliation{
$1.$ Department of Physics, Hong Kong University of Science and
Technology, Clear Water Bay, Hong Kong. \\
$2.$  Laboratoire de physique th\'eorique de la mati\`ere condens\'ee, UMR 7600 C.N.R.S., Universit\'e Pierre et Marie Curie, Paris, France. \\
$3.$  Institut Laue Langevin, BP 156, F-38042 Grenoble Cedex
9, France.
}
\date{\today}

\begin{abstract}
We study the dispersion of the magnons (triplet states) in
SrCu$_2$(BO$_3$)$_2$ including all symmetry-allowed \DM interactions.
We can reduce the complexity of the general Hamiltonian to a 
simpler form   by appropriate rotations of the spin
operators. The resulting Hamiltonian is studied by both
perturbation theory and exact numerical diagonalization on a 32-site
cluster. We argue that the dispersion is dominated by \DM
interactions. We point out which combinations of these anisotropies affect the
dispersion to linear-order, and extract their magnitudes.
\end{abstract}

\pacs{PACS numbers:} \maketitle
\section{Introduction}

Strontium copper boron oxide (SrCu$_2$(BO$_3$)$_2$) is a
two-dimensional antiferromagnet with no long-range magnetic
order.\cite{Kageyama,MiyaharaReview} Heisenberg interactions between
the $S=1/2$ magnetic moments are strong and the geometry is such that
the planes of spin dimers are coupled in a way equivalent to those of
the Shastry-Sutherland model.  \cite{Shastry,Miyahara} This model had
been proposed on purely theoretical grounds, and is remarkable in that
while strongly interacting, it has an {\it exact} dimer ground state,
fully isotropic in spin-space, up to a critical value of the
coupling.\cite{Shastry} It is not integrable, however, and neither the
excited states, nor their dispersions are known exactly. This
motivated many studies of the dynamics of this model.  The existence
of a compound which is well modeled by the Shastry-Sutherland
Hamiltonian, and what is more, is in an intermediate range of
coupling, not far from a quantum critical point, is then of great
interest.  The spin excitations of SrCu$_2$(BO$_3$)$_2$ have been
studied extensively by a variety of experimental techniques (see
\textit{e.g.} [\onlinecite{Kakurai0,Lemmens}] and references below).  An
immediate question is how to extract the couplings and to give a precise
answer to the question: to what extent SrCu$_2$(BO$_3$)$_2$ can be
described by the Shastry-Sutherland model?  \par The first observation
is that while the Shastry-Sutherland model and its ground state are
rotationally invariant, SrCu$_2$(BO$_3$)$_2$ shows departures from spin
isotropy.  Electron spin resonance (ESR),\cite{NojiriESR} far-infrared
spectroscopy of forbidden transitions,\cite{Room1} neutron inelastic
scattering,\cite{Cepas} or latter Raman scattering\cite{Gozar} have
shown the splitting of the spin triplet excitations at
$\mathbf{q}=\mathbf{0}$ and anisotropic behavior with respect to the
direction of an external magnetic field. This was
explained\cite{Cepas} as originating from a \DM interaction (a
correction of spin-orbit origin)\cite{Moriya} whose characteristic
vector is oriented perpendicular to the copper planes, along the
$c$-direction of the crystal. While spin-orbit interactions always
introduce anisotropies at some energy scale, it was a surprise that in
a frustrated model they could dominate the dispersion which generally
arises from larger, isotropic, couplings.  More recently, other
components of the vectors were argued to be required to explain further
experimental findings: another splitting was observed at
$\mathbf{q}=(\pi,0)$,\cite{KakuraiNeutrons1,Gaulin} and was subsequently
associated with in-plane components of the \DM interactions. In
addition, there is an avoided level crossing when the first triplet
state is about to cross the ground state for   magnetic fields 
applied along $c$. This is not compatible with a single-axis
anisotropy along the same direction, and requires additional forms of
anisotropy. The level anti-crossing was argued to be due to the nearest
neighbor \DM interaction,\cite{Miyaharaintra,Kodama,MiyaharaQSS04}
which was  considered as a possible explanation for the X-band ESR
as well.\cite{Zorko} All these components are indeed allowed by the
symmetry of the crystal structure in the low temperature phase of
SrCu$_2$(BO$_3$)$_2$,\cite{Smith,Sparta,Choi} whose copper planes are
slightly buckled. Were it not buckled (as in the high temperature
phase\cite{Sparta}), these components would have vanished. It was for
this reason that they were neglected in the original interpretation of
the neutron inelastic scattering.\cite{Cepas} Nonetheless, because the
symmetry is slightly broken these interactions are expected to be
present and define a smaller energy scale.

A precise model for SrCu$_2$(BO$_3$)$_2$ must therefore include, in
addition to the larger Heisenberg couplings, a complex pattern of \DM
interactions. Since their relative strengths are not precisely known,
it is difficult to establish a hierarchy among them, other than the
probable dominance of the $c$ axis component.

In this paper, we argue, on the basis of exact transformations and
numerical diagonalizations, that we can simplify the model and
determine the parameters of the transformed model by studying the
dispersion of the triplet states.  The dispersion of the triplet
states of the Shastry-Sutherland model, taken alone, is very small,
and this is why smaller interactions are relevant, and even turn out
to be dominant.\cite{Cepas} It is natural to consider \DM interactions
since they are linear in the ratio of the spin-orbit coupling to the
crystal-field splitting, $\lambda$, which is estimated from the
$g$-factor, \textit{i.e.}  $\frac {g-2}{g}$ to be about 0.1. They are,
however, peculiar, in that they are usually expected to have little
effect on the spectrum, at most $\lambda^2$.\cite{rotation} It has
even been claimed that \DM interactions and second-order symmetric
anisotropic couplings conspire to restore the rotational invariance of
the spin excitation spectrum.\cite{rotation} While this can be true
for single-band models in certain geometries, here the frustration of
the lattice makes such arguments inapplicable, as we shall see. We can
in fact classify the \DM components into \textit{reducible} components
(that can be reduced to $\lambda^2$ effects) and \textit{irreducible}
components that have effects on the dispersion linear in the
spin-orbit coupling. The latter thus defines a larger energy scale
which will be the focus of our discussion.

The problem then is to calculate the spin excitation spectrum
reliably. If only isotropic interactions are considered,
SrCu$_2$(BO$_3$)$_2$ is very close in parameter space ($J^{\prime}/J =
0.62$) to a quantum critical point,\cite{Miyahara} ($J^{\prime}/J =
0.68$), which separates the dimer phase (which is known exactly) to a
phase that is not known exactly but may be
quadrumerized.\cite{Koga,Lauechli,Malrieu} Perturbative approaches to
the excitations may therefore turn out to be inaccurate. For instance,
the splitting of the $\mathbf{q}=\mathbf{0}$ triplet energy is given
by $\delta=D^{\prime}_{\perp} g(J^{\prime}/J)$, which is linear in
$D^{\prime}_{\perp}$ and involves a function $g(J^{\prime}/J)$ with
$g(0)=4$, but $g(J^{\prime}/J)=2.0$ for
$J^{\prime}/J=0.62$\cite{Cepas}. This is difficult to calculate by
perturbative techniques because of the proximity with the quantum
critical point. Therefore, in order to explain the dispersion of the
first triplet states in SrCu$_2$(BO$_3$)$_2$, we need not only to
include the relevant \DM interactions, but also to go beyond the
perturbative techniques used earlier for the
dispersion.\cite{Miyahara,Weihong,KnetterDispersion} Indeed, even
carried at high order in perturbation theory, such calculations may
tend to overestimate the dispersion, as we shall see.  For these
reasons, we have carried out an exact numerical study of the model,
including the \DM interactions.

In addition to symmetry-allowed components of the \DM vectors, other
components, that are not allowed by the measured crystal symmetry were
invoked\cite{Jorge,ElShawish} as possible explanations for the
specific heat and the ESR ``forbidden'' transition. We shall not
consider them here because, on one hand, no distortions of the crystal
structure that would allow them have been reported so
far,\cite{Sparta,Choi} and electric-dipole-transitions provide a
symmetry-preserving alternative to explain forbidden
transitions.\cite{Cepas,CepasOptical} Although it is not clear whether
ESR are of magnetic-dipole or electric-dipole nature, it has been
shown that the forbidden transitions observed by far infrared
spectroscopy were clearly electric-dipole.\cite{Room2} Even if the ESR
transitions were magnetic-dipole, they could possibly be explained
by finite-temperature effects.\cite{Miyashita} These components, not
allowed by the symmetry, seem therefore unnecessary at our current
state of knowledge.

Those components which are allowed by symmetry, however, are most
probably important to understand refined experiments, such as the
high-field plateau phases\cite{Onizuka,Takigawa} where typical
condensation energies are much smaller than the isotropic
couplings. For instance, the 1/8 plateau width is of order of 1 Kelvin
and it is questionable whether Heisenberg models alone capture the
essential superstructure of that phase. This provides further motivation
to quantify to what extent the isotropic model applies to
SrCu$_2$(BO$_3$)$_2$, and in what circumstances the anisotropic
couplings can be safely neglected.

The paper is organized as follows. In section \ref{unfrustrated}, we give the
dispersion of the first-triplet excitations for the Heisenberg
Shastry-Sutherland model. In section \ref{model}, we present
the anisotropic model for SrCu$_2$(BO$_3$)$_2$ with \DM interactions,
and in (\ref{mapping}) map it onto a simpler model.  We then calculate the
magnon dispersion both perturbatively (section \ref{perturbation}), and by exact
diagonalization of finite lattices (section \ref{diagonalization}). We discuss
experimental results in section \ref{exp} and extract the couplings that
previously could not be quantified by comparing exact spectra with
experimental results.

\section{Dispersion of the lowest triplet excitation of the Shastry-Sutherland model}
\label{unfrustrated}

We first consider the Shastry-Sutherland model\cite{Shastry} (with no
\DM interaction), as originally used\cite{Miyahara} to
describe SrCu$_2$(BO$_3$)$_2$:
\begin{equation}
H = \sum_{nn} J \textbf{\mbox{S}}_i . \textbf{\mbox{S}}_j  +
\sum_{nnn} J^{\prime} \textbf{\mbox{S}}_i . \textbf{\mbox{S}}_j 
\label{eqn:ssm}
\end{equation}
where $nn$ stands for nearest neighbors (intra-dimer) and $nnn$ for
next nearest neighbors (inter-dimer). The lattice of couplings is as
shown in Figure \ref{Lattice}: nearest neighbour couplings are on the
diagonals in every second square of the lattice of
next-nearest-neighbour couplings (in the real lattice the angles are
such that the diagonals are shorter). The strengths of the Heisenberg
couplings are $J=85$ Kelvin for the nearest-neighbor coupling, and
$J^{\prime}=54$ for the next-nearest-neighbor coupling, as
determined\cite{Miyahara} from susceptibility measurements. An
interplane coupling is present, but it is small ($J^{\prime \prime}=8$
K) and frustrated.\cite{Miyahara} We therefore restrict ourselves to a
two-dimensional version of the model and neglect  $J^{\prime \prime}$. The ratio $J^{\prime}/J \sim 0.62-0.63$
was confirmed to be consistent with the ratio of the first
Raman-active singlet energy to the first triplet energy,\cite{Cepas} 
a measure independent of the magnetic susceptibility. 

\begin{figure}[htbp]
\centerline{
 \psfig{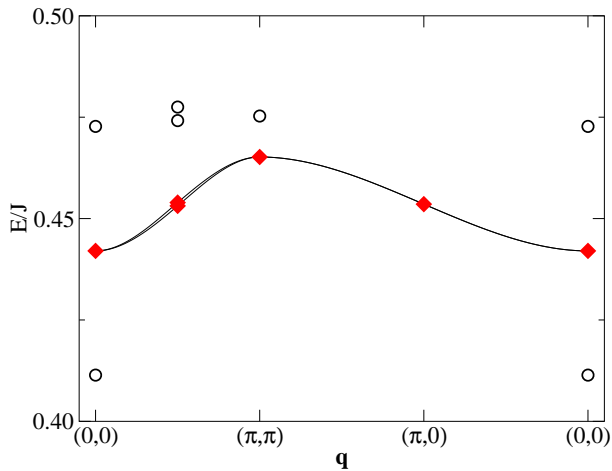}}
\caption{(color online). Dispersion relation of the first triplet
 excitation of the Heisenberg Shastry-Sutherland model (diamonds),
 as calculated by exact numerical diagonalization on a 32-site
 cluster, for $J^{\prime}/J=0.62$. The vertical axis shows the
 excitation energy of those states with respect to the ground state.
The degeneracy of the two triplets
 at ($\pi/2,\pi/2$) is slightly lifted, reflecting the two dimers per
 unit cell. Singlet states nearby are shown as open circles.}
\label{noDM}
\end{figure}

We calculate a few low-lying states of the Shastry-Sutherland model
[Hamiltonian (\ref{eqn:ssm})] on a 32-site cluster using exact
diagonalization.  This cluster is particularly attractive compared to
smaller ones not only because it has smaller finite size effects, but
also because it allows access to the important symmetry points
$(\pi/2,\pi/2)$, $(\pi,\pi)$ and $(\pi,0)$ [\onlinecite{lc2004}]. The
results are given in Fig.~\ref{noDM} for $J^{\prime}/J=0.62$.

The gap between the ground state and the triplet state is $0.442J$ for
the 32-site cluster and essentially independent of the system-size as
shown in table~\ref{bandwidth}. Yet the absence of finite-size effects
on the gap does not mean that all energies have reached their
thermodynamic limit. In fact the dispersion of the 16-site cluster is
much flatter than that of the 20 or 32-site clusters. It has appeared
before (but for the simple square lattice) that the 16-site cluster
can be quite special compared with larger clusters.\cite{SchulzZiman}
Comparison of the bandwidths of the 20-site and 32-site clusters is
also difficult because the reciprocal points are not identical. In
particular the bandwidth of the 20-site cluster is inherently smaller
because of the absence of the $(\pi,\pi)$ reciprocal point: the only
non-zero point in reciprocal space is at
$(\frac{2\pi}{5},\frac{4\pi}{5})$. Therefore while finite-size effects
are definitely smaller than the table~\ref{bandwidth} might suggest it
is not possible to quantify them at present. In absolute values, the bandwidth of
the 32-site cluster remains very small, $\sim 0.023J$, \textit{i.e.}
about $2$ Kelvin. We note that it is smaller than that of the
perturbative calculation,\cite{Weihong,KnetterDispersion} although
this may be explained in part by finite-size effects, but is also
qualitatively different in shape. Consistent with the fact that there
are two dimers per unit-cell, we have in fact not one but two triplet
states which are degenerate at high symmetry points of reciprocal
space. The degeneracies are slightly lifted for intermediate
q-values. This splitting was shown to occur at higher order,
$(J^{\prime}/J)^{10}$ in perturbation theory.\cite{KnetterDispersion}
The gap and especially the dispersion relation are in fact much more
affected by the smaller anisotropic interactions, as will be shown in
the next section.
\begin{table}[htbp]
 \begin{ruledtabular}
 \begin{tabular}{ccc}
cluster size & spin gap& bandwidth\\
\colrule
16 & 0.460\,125$J$ & 0.003\,15$J$ \\
20 & 0.451\,441$J$ & 0.015\,98$J^*$ \\
32 & 0.442\,026$J$ & 0.023\,16$J$ \\
 \end{tabular}
 \end{ruledtabular}
\caption{ The spin gap and bandwidth of the first triplet excitation
  in the Shastry-Sutherland model for different cluster sizes. $^*$The
  bandwidth of the 20-site cluster cannot be directly compared with
  the others because of the absence of the $(\pi,\pi)$ reciprocal point.}
\label{bandwidth}
\end{table}

\section{Model and exact mapping }

\subsection{The general anisotropic model}
\label{model}

\begin{figure}[htbp]
\centerline{
 \psfig{file=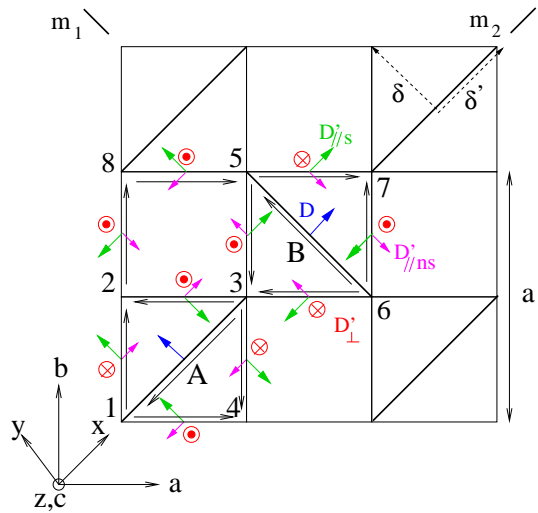,width=7cm,angle=0}}
\caption{(color online). Shastry-Sutherland lattice with
 Dzyaloshinskii-Moriya interactions allowed by the symmetries of
 SrCu$_2$(BO$_3$)$_2$. Colored (short gray) arrows are the components
 of the \DM vectors for each bond (long arrows show the prescription
 for the order of the operators, a black arrow from $i$ to $j$
 indicates that we should take $ \textbf{\mbox{D}} \cdot (
 \textbf{\mbox{S}}_i \times \textbf{\mbox{S}}_j )$ with the given $
 \textbf{\mbox{D}}$). $m_1$ and $m_2$ are mirror planes, but the
 $(ab)$ plane is not a mirror plane in the low-temperature phase. This
 allows for all in-plane components of $ \textbf{\mbox{D}}$. }
\label{Lattice}
\end{figure}

We now consider a model with \DM interactions, based on the symmetries
of the low temperature phase.\cite{Smith,Sparta,Choi} The model
contains isotropic couplings,\cite{Miyahara} and both
inter-dimer\cite{Cepas,KakuraiNeutrons1} and
intra-dimer \DM interactions.\cite{Miyaharaintra} It
reads:
\begin{eqnarray}
H &=& \sum_{nn} \left[ J \textbf{\mbox{S}}_i . \textbf{\mbox{S}}_j  + \textbf{\mbox{D}}_{ij}. ( \textbf{\mbox{S}}_i \times \textbf{\mbox{S}}_j) \right] \nonumber  \\ &+& \sum_{nnn} \left[J^{\prime} \textbf{\mbox{S}}_i . \textbf{\mbox{S}}_j  + \textbf{\mbox{D}}_{ij}^{\prime} . ( \textbf{\mbox{S}}_i \times \textbf{\mbox{S}}_j) \right]
\label{ham}
\end{eqnarray}
where $nn$ stands for nearest neighbors (intra-dimer) and $nnn$ for
next-nearest neighbors (inter-dimer). The different components of the
vectors $\textbf{\mbox{D}}_{ij}$ and $\textbf{\mbox{D}}_{ij}^{\prime}$
are given in Fig.~\ref{Lattice}.
 They are obtained by using 
transformation properties of pseudo-vectors and the symmetries of the
crystal structure.\cite{Moriya} For instance, when a bond connecting two
spins belongs to a mirror plane, the \DM vector must be perpendicular
to it. In the high temperature phase ($T>400$ K),\cite{Sparta} the
$(ab)$ plane is a mirror plane and the allowed components (only the
inter-dimer ones) must therefore be parallel to the crystallographic
$c$-axis. This is the $\mathbf{D}^{\prime}_{ij}=D_{\perp}^{\prime}
\hat{\mathbf{e}}_z$ ($\hat{\mathbf{e}}_z$ is the unit vector along
$z=c$) we considered before. In the low temperature phase, the $(ab)$
plane is slightly buckled, and some other in-plane components are allowed,
but are expected to be smaller.\cite{KakuraiNeutrons1} The remaining
$m_1$, $m_2$ mirror planes allow us to determine the complete pattern of
\DM vectors. In-plane components of $\mathbf{D}^{\prime}_{ij}$ can be
decomposed into staggered $D^{\prime}_{\parallel,s}$ and non-staggered
$D^{\prime}_{\parallel,ns}$ orthogonal components. We obtain
\begin{equation}
\mathbf{D}^{\prime}_{ij} = D^{\prime}_{\parallel,s} \hat{\mathbf{e}}_{s,ij} + D^{\prime}_{\parallel,ns} \hat{\mathbf{e}}_{ns,ij} + D_{\perp}^{\prime} \hat{\mathbf{e}}_z
\end{equation}
where the unit vectors are, respectively, $\hat{\mathbf{e}}_{s,ij}
\equiv \pm \hat{\mathbf{e}}_y$ and $\hat{\mathbf{e}}_{ns,ij} \equiv
\pm \hat{\mathbf{e}}_x$ (for dimers A) or $\hat{\mathbf{e}}_{s,ij} \equiv
\pm \hat{\mathbf{e}}_x$ and $\hat{\mathbf{e}}_{ns,ij} \equiv \pm
\hat{\mathbf{e}}_y$ (dimers B) (Fig.~\ref{Lattice}).  The intra-dimer
vector $\mathbf{D}_{ij}$ has only in-plane components,
\begin{equation}
\mathbf{D}_{ij} = D \hat{\mathbf{e}}_{ij}
\label{Ddef}
\end{equation}
with the unit-vector $\hat{\mathbf{e}}_{ij}=\hat{\mathbf{e}}_y$
(dimers A) or $\hat{\mathbf{e}}_{ij}=\hat{\mathbf{e}}_x$ (dimers B).
Note that because the interaction is a vector product of spins, we
have to be careful with the order of the operators in the
definitions. The order is defined in Fig.~\ref{Lattice} for each bond
by the black arrows. For instance, for the bond $1 \rightarrow 2$, we
define $\mathbf{D}^{\prime}_{12}=D^{\prime}_{\parallel,s}
\hat{\mathbf{e}}_y + D^{\prime}_{\parallel,ns} \hat{\mathbf{e}}_x
-D^{\prime}_{\perp} \hat{\mathbf{e}}_z $. The strengths of three of
these four couplings, $D$, $D_{\parallel,s}^{\prime}$,
$D_{\parallel,ns}^{\prime}$, are unknown. In the next paragraph, we
show how to simplify the Hamiltonian (\ref{ham}) such as to eliminate
two of these couplings. Eventually only one parameter remains to be
determined and we will determine it with the help of the magnon
dispersion at finite wave-vector.

\subsection{Mapping}
\label{mapping}
We map the Hamiltonian (\ref{ham}) onto a simpler
Hamiltonian with no intra-dimer and no uniform \DM interactions, by
means of rotations\cite{rotation} of spin operators. Contrary to
previous situations where this approach has been applied,\cite{rotation,Oshikawa} it turns out here that we cannot
eliminate all components, because of the frustration of the lattice but we can
nonetheless simplify the model. For this, we perform two sorts of
rotations.  First, we rotate the two spin
operators of the dimers in opposite directions, and apply the same
operation to all unit cells. Second, we rotate the spin operators from
dimer to dimer by applying the same operation to the two spin
operators of the same dimer.  The first transformation uses local
rotations (the same for all unit-cells) for the two spins $i$ and $j$
of the dimer $i,j$, that are defined by: 
\begin{eqnarray}
  \textbf{\mbox{S}}_i^{\prime} = {\cal R}(\theta/2,\hat{\textbf{\mbox{e}}}_{ij}) \textbf{\mbox{S}}_i \hspace{5cm} \\
= (1-\cos \frac{ \theta}{2}) [\hat{\textbf{\mbox{e}}}_{ij}.\textbf{\mbox{S}}_i] \hat{\textbf{\mbox{e}}}_{ij} + \cos\frac{ \theta}{2} \textbf{\mbox{S}}_i - \sin \frac{ \theta}{2} \textbf{\mbox{S}}_i \times \hat{\textbf{\mbox{e}}}_{ij} 
\label{transformation} 
\end{eqnarray}
with $\tan \theta=D/J$,\cite{rotation} where $D$ and
$\hat{\mathbf{e}}_{ij}$ are defined by (\ref{Ddef}). The second spin
$j$ of the same dimer is rotated in the opposite sense ($\theta
\rightarrow - \theta$). For the second dimer of the unit-cell, the
spins are rotated similarly, about an axis that is perpendicular to
that, according to the local $\textbf{\mbox{D}}_{ij}$ (see
Fig.~\ref{Lattice}). 

We now restrict consideration to terms that are linear in the \DM coupling strength
in the transformation (\ref{transformation}), because we are looking
for linear effects in the dispersion of excitations. In fact the
transformation (\ref{transformation}) does produce some anisotropic
exchange of order $D^2/J$,\cite{rotation} which we neglect because (i)
we have not taken into account real anisotropic couplings which occur
at the same order of magnitude, and more importantly, (ii) the
spectrum is fully dominated by effects that are linear in $D$, as we
shall show.

In terms of the primed spins, the Hamiltonian (\ref{ham}) is mapped
onto (up to terms of order $D^2/J$):
\begin{equation}
H = \sum_{nn} J \textbf{\mbox{S}}_i^{\prime} . \textbf{\mbox{S}}_j^{\prime}  + \sum_{nnn} \left[J^{\prime} \textbf{\mbox{S}}_i^{\prime} . \textbf{\mbox{S}}_j^{\prime}  + \textbf{\mbox{D}}_{ij}^{\prime} . ( \textbf{\mbox{S}}_i^{\prime} \times \textbf{\mbox{S}}_j^{\prime}) \right]
\label{hammapped}
\end{equation}
where the $\textbf{\mbox{D}}_{ij}^{\prime}$ are now modified: the
staggered $c$-component is unchanged, whereas the
strengths of the in-plane components (either staggered or uniform) are
modified according to
\begin{eqnarray}
 D^{\prime}_{\parallel,ns} &\rightarrow& D^{\prime}_{\parallel,ns} + \frac{J^{\prime}}{2J} D \label{newDMns} \\
 D^{\prime}_{\parallel,s} &\rightarrow& D^{\prime}_{\parallel,s} + \frac{J^{\prime}}{2J} D \label{newDMs} \\
D^{\prime}_{\perp} &\rightarrow& D^{\prime}_{\perp} \label{newDM}
\end{eqnarray}
This is an exact transformation that is valid at all order in
$J^{\prime}/J$. As a second step, we now proceed to eliminate the
non-staggered component of $ \textbf{\mbox{D}}_{ij}^{\prime}$, which
is defined by its strength $D_{\parallel,ns}^{\prime}$
(eq. (\ref{newDMns})) and a vector direction either along $x$ or $y$
(see Fig.~\ref{Lattice}). This is made possible because the
$D_{\parallel,ns}^{\prime}$ sum up to zero along any closed loop of
the lattice. When going around closed loops, the rotation angles are
fixed by the \DM strength and sign; therefore there is a compatibility
condition for the last spin.\cite{Oshikawaprivate} Consider for
instance the closed loop $(2-8-5-3-2)$ in Fig.~\ref{Lattice}, the sum
of the $D_{\parallel,ns}^{\prime}$ is zero along this loop. Note, however,
that the sum of the $D_{\perp}^{\prime}$ does not vanish, nor does
the sum of the $D_{\parallel,s}^{\prime}$ along $(5-7-6-5)$, for
instance, so these cannot be eliminated. We now rotate the operators
accordingly, and restrict to the first-order terms in
$D_{\parallel,ns}^{\prime}$:
\begin{equation}
\tilde{\textbf{\mbox{S}}}_i =  {\cal R}(\theta_i, \hat{\textbf{\mbox{e}}}_{y}) {\cal R}(\psi_i, \hat{\textbf{\mbox{e}}}_{x}) \textbf{\mbox{S}}_i^{\prime} 
\label{transformation2}
\end{equation}
In this expression, $\theta_i =  (D^{\prime}_{\parallel,ns}/J^{\prime}) \textbf{\mbox{R}}_i \cdot \hat{\textbf{\mbox{e}}}_x$, and $\psi_i =  (D^{\prime}_{\parallel,ns}/J^{\prime}) \textbf{\mbox{R}}_i \cdot \hat{\textbf{\mbox{e}}}_y$, where  $\hat{\textbf{\mbox{e}}}_{x,y}$ are unit vectors along the $x$ and $y$ directions. $\textbf{\mbox{R}}_i$ is the position of the dimer to which the spin $i$ belongs, in units of the inter-dimer spacing. The two spins of the same dimer are rotated in the same way, so as not to generate other intra-dimer \DM interactions. The final Hamiltonian becomes:
\begin{eqnarray}
&& \hspace{-0.7cm} H = \sum_{nn} J \tilde{\textbf{\mbox{S}}}_i . \tilde{\textbf{\mbox{S}}}_j  + \sum_{nnn} \left[J^{\prime} \tilde{\textbf{\mbox{S}}}_i . \tilde{\textbf{\mbox{S}}}_j  + \tilde{\textbf{\mbox{D}}}_{ij}
 . ( \tilde{\textbf{\mbox{S}}}_i \times \tilde{\textbf{\mbox{S}}}_j) \right] 
\label{hammapped1} \\
&& \tilde{\textbf{\mbox{D}}}_{ij} = \tilde{D}_{\parallel} \hat{\mathbf{e}}_{s,ij} + \tilde{D}_{\perp} \hat{\mathbf{e}}_{z} \label{remains} \\
&& \tilde{D}_{\parallel} = D_{\parallel,s}^{\prime} + \frac{J^{\prime}}{2J} D  \hspace{1cm}
 \tilde{D}_{\perp} = D_{\perp}^{\prime} \label{newD} 
\end{eqnarray}
where $\hat{\mathbf{e}}_{s,ij}$ is the unit-vector along the
inter-dimer staggered \DM vector, defined in Fig.~\ref{Lattice}.  The
transformed model (\ref{hammapped1}) is simpler and can be used instead of the
original Hamiltonian (\ref{ham}), as far as energies are concerned.
The remaining \DM components (\ref{remains}) which we call
\textit{irreducible} cannot be eliminated by rotations. Geometrically this is
because they do not sum to zero when turning around closed loops of
the Shastry-Sutherland lattice. As a consequence, linear effects are
present in the spectrum.

We shall proceed with the Hamiltonian (\ref{hammapped1}) since the two
Hamiltonians have exactly the same energy spectrum up to terms of
order $D^2/J$, $D_{\parallel,ns}^{\prime 2}/J$. Note that this is
valid only at zero magnetic-field: at finite fields along $z$, for
instance, the first transformation creates a staggered transverse
field\cite{Oshikawa} (which must already be present according to the
local environment surrounding the copper ions\cite{Miyaharaintra}) and
the second transformation creates rotating fields.

\section{Perturbative calculation of the excitation spectrum}
\label{perturbation}

We calculate the magnon dispersion and dynamical
structure factor of model (\ref{hammapped1}), by first-order
perturbation theory in the \DM couplings and
$J^{\prime}/J$.\cite{Cepas,KakuraiNeutrons1} Since the latter is not
small for SrCu$_2$(BO$_3$)$_2$ ($J^{\prime}/J=0.62 - 0.63$), we will
supplement this approximation with exact numerical results in 
section \ref{diagonalization} but perturbation theory already gives some qualitative
insights into the problem.

\begin{figure}[htbp]
\centerline{
 \psfig{file=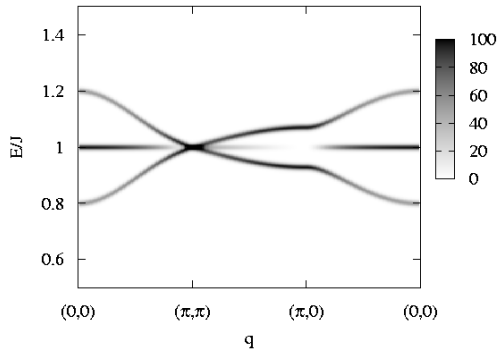,width=10cm,angle=-0}}
\caption{($J^{\prime}/J \rightarrow 0$) Dispersion relation and
dynamical structure factor (grey scale) of the first triplet excitations in
the limit of small $J^{\prime}/J$. Note the overall dimer form
factor $M_q$ is not included in the picture. For $(\pi,0)$, the middle mode
has no intensity. For the clarity of the picture, $\tilde{D}_{\perp}=0.1J$, and $\tilde{D}_{\parallel} = 0.05J$ were used.}
\label{noJ}
\end{figure}

We start with the exact dimer ground state, which is a product of
singlets on each dimer. We consider the following one-particle excited
states, which consist of promoting a dimer into a triplet state with
$S^z=0, \pm 1$ quantum number. They are the eigenstates for zero
$J^{\prime}$ and $D$, and we use first-order perturbation theory in
$J^{\prime}/J$, $\tilde{D}_{\parallel}$ and
$\tilde{D}_{\perp}$. Because there are two dimers per unit-cell, there
are two such triplet states depending on which dimer A or B is in the
triplet state. When $\tilde{D}_{\perp}$ is added, the total rotation
invariance is broken but $S^z$ is still a good quantum number thanks
to the rotation about $c$. In this case, the eigenstates within this
subspace of states are given by:\cite{Cepas}
\begin{equation}
| S^z, \mathbf{q} , \pm \rangle = \frac{1}{\sqrt{2}} \left(|S^z,\mathbf{q},A \rangle \pm i |S^z, \mathbf{q},B \rangle \right) 
\label{0states}
\end{equation}
with $S^z = 0,\pm 1$ the $z$ component of the total spin of a dimer, and $|S^z,\mathbf{q},A \rangle$ is the Fourier transform of the triplet state on the dimer A of unit-cell $i$, $|S^z,i,A \rangle $.  Now in addition we consider the effect of inter-dimer
staggered components of the \DM vectors \textit{i.e.} model
(\ref{hammapped1}).  $\tilde{D}_{\parallel}$ breaks the rotation
invariance about $c$ and therefore mixes various $S^z$
components of (\ref{0states}). The Hamiltonian within the subspace of (\ref{0states})
splits into two blocks. The first block for   
 $(|+1,\mathbf{q}, + \rangle,|0,\mathbf{q}, -
\rangle,|-1,\mathbf{q}, +  \rangle) $ reads
\begin{equation}
\left( \begin{array}{ccc}
J+\tilde{D}_{\perp} f(\mathbf{q}) & -i \tilde{D}_{\parallel} g(\mathbf{q}) & 0 \\
 i \tilde{D}_{\parallel} g(\mathbf{q})^* & J &  i \tilde{D}_{\parallel} g(\mathbf{q}) \\
0 &  -i \tilde{D}_{\parallel} g(\mathbf{q})^* & J-\tilde{D}_{\perp} f(\mathbf{q}) \end{array} \right)
\end{equation}
where we have $f(\mathbf{q}) =2 \cos (q_a /2) \cos (q_b /2)$ and
$g(\mathbf{q})=-\left(\sin [(q_a+q_b)/2] + i \sin [(q_a-q_b)/2]
\right)/\sqrt{2} $, $\mathbf{q}=(q_a,q_b)$. The lattice spacing is
taken such that $a=1$. The second block for the three other states, is
identical up to the change $(\tilde{D}_{\parallel},\tilde{D}_{\perp})
\rightarrow -(\tilde{D}_{\parallel},\tilde{D}_{\perp})$.  The six
resulting triplet states are denoted by $|t^m_{\mathbf{q}} \rangle$
and have energies (twice degenerate because of identical blocks):
\begin{eqnarray}
E^{\pm}(\mathbf{q}) &=& J \pm \sqrt{\tilde{D}_{\perp}^{2} f(\mathbf{q})^2 + 2 \tilde{D}_{\parallel}^2 |g(\mathbf{q})|^2} \equiv J \pm \frac{\delta_{\mathbf{q}}}{2} \nonumber \\
E^0(\mathbf{q}) &=& J \label{d}
\end{eqnarray}
The dispersion relations of ``triplet'' states (\ref{d}) are given in
Fig.~\ref{noJ}. They are valid only at first-order in $J^{\prime}/J$
and higher-order corrections are not included here.  We shall see
below how the overall shape and degeneracies are modified when one
goes beyond perturbation theory. For the moment, this defines two
splittings at $\mathbf{q}=(0,0)$ and $\mathbf{q}=(\pi,0)$ that are given by:
\begin{eqnarray}
\delta_{(0,0)} &=& 4\tilde{D}_{\perp}=4D_{\perp}^{\prime} \label{delta00} \\
\delta_{(\pi,0)}&= & 2\sqrt{2} \tilde{D}_{\parallel} = 2\sqrt{2} \left(D_{\parallel,s}^{\prime} + \frac{J^{\prime}}{2J} D \right)  \label{deltapi0}
\end{eqnarray}
where the right-hand sides give the expressions of the splittings in
terms of the original couplings of Fig.~\ref{Lattice}, according to
the transformations (\ref{newDMs}) and (\ref{newDM}). It is
interesting to note that measuring these splittings at different
wave-vectors allows, in principle, separation of the two
couplings. This is because of the different staggering patterns of the
\DM vectors in space, which is itself a consequence of the crystal
symmetry.

In order to compare with neutron inelastic scattering at finite
wave-vectors, we have computed the neutron cross-section, given
by\cite{Squires}
\begin{eqnarray}
\left( \frac{d^2 \sigma }{d \omega d \Omega}\right) \propto \sum_m S^m(\mathbf{q}) \delta \left(\omega - E^m(\mathbf{q}) \right) \\
S^m(\mathbf{q}) =  \sum_{\alpha \beta} \left(\delta_{\alpha \beta} - \frac{\mathbf{q}_{\alpha} \mathbf{q}_{\beta}}{\mathbf{q}^2} \right) \langle 0 |  S_{\textbf{\mbox{q}}}^{\alpha} | t^m_{ \textbf{\mbox{q}}} \rangle \langle t^m_{ \textbf{\mbox{q}}} |  S_{\textbf{-\mbox{q}}}^{\beta} | 0 \rangle
\label{int}
\end{eqnarray}
Note that $S_{\mathbf{q}}^{\alpha}$ is, strictly speaking, the real
spin operator, not that obtained after the transformations are
performed, $\tilde{S}_{\mathbf{q}}^{\alpha}$. We can in fact replace
$S_{\mathbf{q}}^{\alpha}$ by $\tilde{S}_{\mathbf{q}}^{\alpha}$ because
it implies small intensity corrections, of order $D/J$, which we shall
discuss later in the paper. The cross-section involves calculating all
matrix elements. Some do not contribute for special directions of the
reciprocal vector, such as for those along the reciprocal line
$\mathbf{q}=(q,0)$ ($\mathbf{q}$ parallel to $a$ gives geometrical
factors $q_a^2/q^2=1$, $q_b^2=q_c^2=0$), and only $\langle
t^m_{\textbf{\mbox{q}}} | S_{\textbf{\mbox{q}}}^{b} | 0 \rangle $ and
$ \langle t^m_{ \textbf{\mbox{q}}} | S_{\textbf{\mbox{q}}}^{c} | 0
\rangle$ contribute. For $\mathbf{q}=(q,q)$ and $\mathbf{q}=(q,0)$,
the intensities have the form
\begin{eqnarray}
S^0(\mathbf{q}) &=& \left( \frac{2 \tilde{D}_{\perp} f(\mathbf{q})}{\delta_{\mathbf{q}}}  \right)^2 M_{\mathbf{q}} \label{I01}  + \cdots \\
S^{\pm}(\mathbf{q}) &=&\left[  \frac{1}{2} + \left(\frac{2 \tilde{D}_{\parallel} |g(\mathbf{q})|}{\delta_{\mathbf{q}}
 } \right)^2 \right] M_{\mathbf{q}}  + \cdots \label{I+1}
\end{eqnarray}
where $M_{\mathbf{q}}= \sin^2 \mathbf{q} \cdot \mathbf{\delta} +
\sin^2 \mathbf{q} \cdot \mathbf{\delta}^{\prime} $ is the form factor
of the unit-cell, $2\delta$ and $2\delta^{\prime}$ being the vectors
along the dimers A and B (Fig.~\ref{Lattice}). The dots represent 
corrections of order ${\cal O}(D,D_{\parallel,ns}^{\prime})$, whereas
the intensities are ${\cal O}(1)$. We will come back to this when
discussing the effect of $D$ and $D_{\parallel,ns}^{\prime}$.  For
$\mathbf{q}=(\pi,q)$, the $q$ dependence comes only from the
geometrical factor:
\begin{eqnarray}
S^0((\pi,q)) &=& \frac{q^2 }{\pi^2 + q^2 } M_q + \cdots \label{I02} \\
S^{\pm}((\pi,q)) &=&  \frac{1}{2} \frac{2\pi^2 + q^2 }{\pi^2 + q^2 } M_q + \cdots \label{I+2} 
\end{eqnarray}
The intensities (eqs. (\ref{I01})-(\ref{I+2})) are represented in grey
scale in Fig.~\ref{noJ} ($M_q$ is not included).  We see, in particular, that the intensity of
the middle mode at $(\pi,q=0)$ is zero.  This is consistent with the
presence of only two modes in neutron scattering experiments at that
wave-vector,\cite{KakuraiNeutrons1,Gaulin} and allows us to identify
these modes as the lower and upper modes (as will be confirmed by
their energies later as well).

\section{Exact numerical diagonalization}
\label{diagonalization}

\begin{figure}[htbp]
\centerline{}
\vspace{0.1cm}
\centerline{
 \psfig{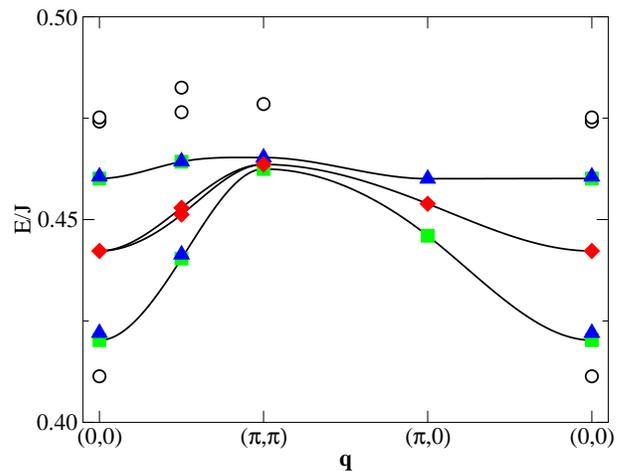} }
\centerline{}
\vspace{0.1cm}
\caption{(color online) 
Same as Fig.~\ref{noDM} but for the Hamiltonian (\ref{hammapped1}) with
$J'/J=0.62$, $\tilde{D}_{\perp}=0.02J$ and 
$\tilde{D}_{\parallel}=0.005J$. Lines are guides
to the eye and join together states which would correspond to $S^z=+1$
(triangles), $-1$ (squares), and 0 (diamonds) in the
limit $\tilde{D}_{\parallel}\rightarrow 0$. Note that some of them are indistinguishable for the upper mode.}
\label{DM}
\end{figure}

We calculate a few low lying states of the Hamiltonian
(\ref{hammapped1}) on the 32-site cluster using exact
diagonalization. Note that the $\tilde{D}_{\parallel}$ term in
(\ref{hammapped1}) breaks spin rotational symmetry and the dimension
of the Hilbert space now becomes $\sim5\times10^8$, as compare to
$\sim7\times10^7$ for the 32-site cluster of the Shastry-Sutherland
model. Results are shown in Fig.~\ref{DM}. The parameters are
$J'/J=0.62$, $\tilde{D}_{\perp}=0.02J$ and
$\tilde{D}_{\parallel}=0.005J$. It is to be compared with
Fig.~\ref{noDM} which corresponds to $\tilde{D}_{\perp}=
\tilde{D}_{\parallel}=0$. Since $S^z$ is no longer a good quantum
number, degeneracies in those triplet states in Fig.~\ref{noDM} are
lifted. Nevertheless, they can still be grouped together according to
whether their energies increase, decrease or remain unchanged in the
presence of a weak magnetic field along the $c$ direction, which
correspond to $S^z=+1$, $-1$, and 0 respectively in the limit
$\tilde{D}_{\parallel} \rightarrow 0$. In Fig.~\ref{DM} states
belonging to the same group at different {\bf q} are joined together
by a line. Note that the middle state, which corresponds to the
$S^z=0$ and indicated by diamonds, is almost unchanged with respect
to Fig.~\ref{noDM}. But the others acquire a larger dispersion that in
any case remains small compared to the gap.
\begin{figure}[htbp]
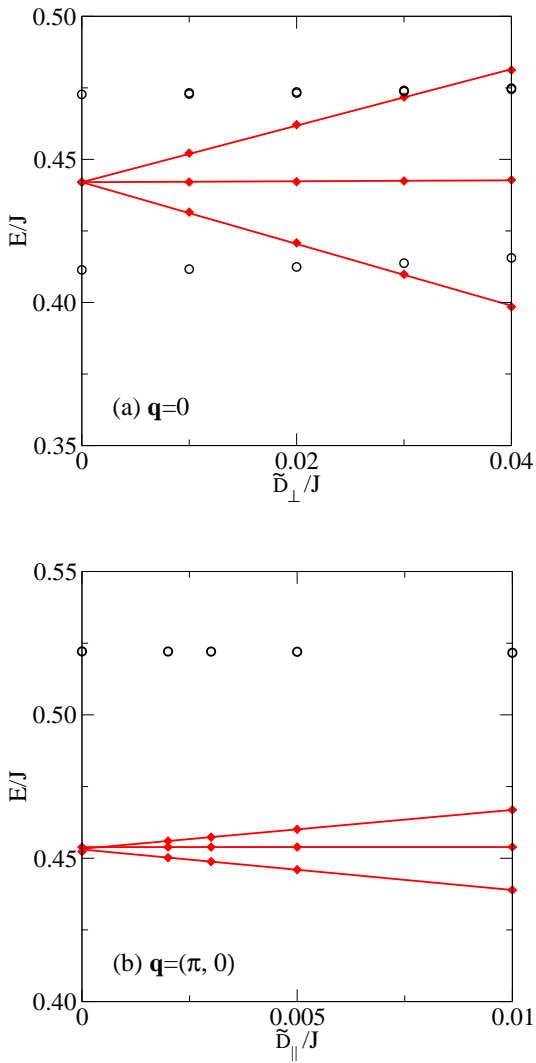

\centerline{
 \psfig{file=Fig/q0.eps,width=7cm,angle=0}}
\centerline{}
\centerline{}
\centerline{ \psfig{file=Fig/qpi.eps,width=7cm,angle=0}}
\caption{
(a) Effect of $\tilde{D}_{\perp}$ on the
$\mathbf{q}=\mathbf{0}$ triplet (diamonds) and singlet
energies (open circles) when $\tilde{D}_{\parallel}=0$, and (b)
effect of $\tilde{D}_{\parallel}$ on the $\mathbf{q}=(\pi,0)$ states
when $\tilde{D}_{\perp}=0.02J$. Numbers are calculated using exact
diagonalization on the 32-site cluster. Lines are linear fittings to
triplet energies.
}
\label{D}
\end{figure}
Next we study in detail the effects of $\tilde{D}_{\perp}$ and
$\tilde{D}_{\parallel}$ on the splitting of the first triplet state.
Fig.~\ref{D}(a) shows the splitting in the $\mathbf{q}=\mathbf{0}$
``triplet'' state at various $\tilde{D}_{\perp}$ and
$\tilde{D}_{\parallel}=0$. We found that a small
$\tilde{D}_{\parallel}$ ($\sim0.005J$) has no noticeable effect on the
splitting of the $\mathbf{q}=\mathbf{0}$ states. Therefore
Fig.~\ref{D}(a) can be regarded as showing the effect of
$\tilde{D}_{\perp}$ on the splitting of those states.  Fig.~\ref{D}(b)
shows the splitting of the first triplet state at $\mathbf{q}=(\pi,0)$
at $\tilde{D}_{\perp}=0.02J$ and various $\tilde{D}_{\parallel}$.  It
is clear that the effect of $\tilde{D}_{\parallel}$ is to modify the
dispersion around the reciprocal point $(\pi,0)$.  At $(\pi,0)$,
Fig.~\ref{D}(b) shows that the energies change linearly as function
of the coupling, except at very small $\tilde{D}_{\parallel}$. In this
region, a very small gap already exists, and is given by
$\tilde{D}_{\perp} (J^{\prime}/J)^{\alpha}$ (where $\alpha > 1$). In
absolute values we have found a gap of $1.6\times10^{-3}J$ (for
$\tilde{D}_{\perp}=0.02J$).  The tiny gap is amplified by
$\tilde{D}_{\parallel}$ and the effect is perturbative in the strength
of the \DM interactions. Note that the slopes, however, are a function
of $J^{\prime}/J$ and are non-perturbative in this coupling.  First,
for values of $J^{\prime}/J$ close to the critical point, the
separation of the components as in (\ref{delta00}) and
(\ref{deltapi0}) is no longer exactly true, but the effect of
$\tilde{D}_{\parallel}$ at $\mathbf{q}=\mathbf{0}$ remains small (and
conversely) in the regime of interest. Second, the prefactors of the
splittings between the upper and lower modes are modified compared to
the perturbative results.  From linear fittings as shown in
Fig.~\ref{D}, the splitting at $\mathbf{q}=\mathbf{0}$ is found to be
$\sim2.08\tilde{D}_{\perp}$ instead of $4 \tilde{D}_{\perp}$ in
(\ref{delta00}).  This compares very favorably with a previous
calculation using the 20-site cluster,\cite{Cepas} thus showing that
size effects are small.  Similarly at $\mathbf{q}=(\pi,0)$ the
splitting is $\sim2.80\tilde{D}_{\parallel}$ instead of $2 \sqrt{2}
\tilde{D}_{\parallel}$ in (\ref{deltapi0}).  Note that the
renormalization at $\mathbf{q}=(\pi,0)$ is in fact very small compared
to that at $\mathbf{q}=\mathbf{0}$, and solves the puzzle previously
noticed,\cite{Gaulin} as we shall explain below.

The spectra show other minor features: at $\mathbf{q}=\mathbf{0}$, for
instance, the lowest state is split into two (Fig.~\ref{DM}). This is
to be expected  when there are several anisotropy axes  and  no Kramers degeneracy in a system with
an even number of spins per unit-cell. We can simply understand
these additional splittings by perturbation theory: couplings with
higher energy states open gaps which scale as
$\tilde{D}_{\parallel}^{2}$ (which we verified by exact
diagonalization). The exact calculation shows other $\lambda^2$
effects. 
We cannot make quantitative comparison of    effects at this order
to real experiments
because anisotropic super-exchange is also present at order
$\lambda^2$ and are neglected in the present work. $\lambda^2$ effects may
perhaps be accessible to ESR at $\mathbf{q}=\mathbf{0}$, but it might
be difficult to disentangle the second-order effects of \DM from that
of real asymmetric couplings.

\section{Extracting the couplings from experiments}
\label{exp}

\begin{figure}[htbp]
\vspace{0.65cm}
\centerline{
 \psfig{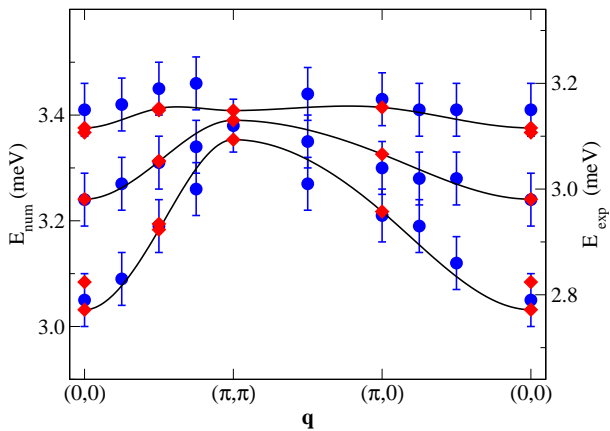} }
\caption{(color online). Magnon dispersion : exact diagonalization with \DM interactions (diamonds) and experimental data  from Ref.~[\onlinecite{KakuraiNeutrons1}] (circles). Lines are guides to the eyes. Parameters are $\tilde{D}_{\perp} = 0.18$ meV, and $\tilde{D}_{\parallel} = 0.07$ meV. $J^{\prime}/J$ was kept equal to $0.62$. This is responsible for the shift in the overall scale
of the calculated energies. }
\label{DMexp}
\end{figure}

\textit{Triplet states.} We now use these results to extract the
anisotropic couplings from neutron inelastic scattering experiments.
At $\mathbf{q}=\mathbf{0}$, we have seen that the energy is unchanged
between the 20-site cluster\cite{Cepas} and the present 32-site
cluster, so the estimation of $D_{\perp}^{\prime}=0.18$ meV is
unmodified.  Regarding the intensity at $\mathbf{q}=\mathbf{0}$,
earlier perturbative calculation\cite{Cepas} compares well with exact
numerical diagonalization on a 20-site cluster,\cite{ElShawish1} so we
have not carried out intensity calculations on the 32-site problem,
but will rely on perturbative results.  At $\mathbf{q}=(\pi,0)$, we
have seen indeed that the intensity of the middle mode is zero
(section \ref{perturbation}).  This is consistent with the observation
of vanishing intensity for the middle mode in neutron
scattering.\cite{KakuraiNeutrons1,Gaulin} The splitting seen at
$\mathbf{q}=(\pi,0)$ can therefore be interpreted as the gap between
the lower and upper state. As we have seen, the splitting at
$\mathbf{q}=(\pi,0)$ is given by $\tilde{D}_{\parallel}
f(J^{\prime}/J)$, with $f(0)=2\sqrt{2}$ and $f(0.62) = 2.80$ (which is
in fact within 1 \% of the perturbative result). This allows to
extract $\tilde{D}_{\parallel} \sim 0.07$ meV, taking a splitting
$\delta(\pi,0)=0.2 $ meV \cite{KakuraiNeutrons1,Gaulin}. This is the
value we extracted before relying on the perturbative
expression.\cite{KakuraiNeutrons1} We have here calculated the
renormalization coefficient at $(\pi,0)$, and shown that that
coefficient is quite different from that at $(0,0)$; therefore
applying the same renormalization to both splittings\cite{Gaulin}
leads to an artificially large in-plane \DM coupling.  This explains
why the ratio appeared so large previously. In fact, the correct ratio
is $\tilde{D}_{\parallel}/\tilde{D}_{\perp}\simeq0.4$ with
$D_{\perp}^{\prime} = 0.18$ meV. We recall that
$\tilde{D}_{\parallel}$ is a linear combination of two interactions
(see below), and as such may explain why it is larger than expected on
the basis of the small buckling of the crystal structure. We therefore
conclude that from the splittings at $\mathbf{q}=(0,0)$ and $\mathbf{q}=(\pi,0)$, we
constrain the couplings:
\begin{eqnarray}
D_{\perp}^{\prime} = 0.18~\mbox{meV} \\
D_{\parallel,s}^{\prime}+ \frac{J^{\prime}}{2J} D  = 0.07~\mbox{meV}
\end{eqnarray}
These values also explain the dispersion of the
excitations as shown in Fig.~\ref{DMexp}, at least for the additional reciprocal
points available in the 32-site cluster (diamonds).
Note that the scales of experiment and numerical calculations 
are slightly shifted. The shift could be eliminated by
an optimized choice of $J^{\prime}/J$ in the range 0.62-0.63,
consistent with previous estimates.

How could we, at least in principle, determine individually $D$,
$D_{\parallel,s}^{\prime}$ and $D_{\parallel,ns}^{\prime}$? As we have
shown, the effect of \DM on the spectrum is large only for the two
effective components that could not be eliminated by our
transformations. To determine the other components, the effect of an
external magnetic field may provide more constraints. However,
staggered gyromagnetic tensors come into play, that are also unknown.
Alternatively, we can use neutron scattering
\textit{intensities}. First, we calculate the correction to the
expressions of the intensities (eq. (\ref{I01}) or (\ref{I02})) at $\mathbf{q}=(\pi,0)$,
coming from the intra-dimer \DM interaction, $D$. Indeed the spin
operators $S_{\mathbf{q}}^{\alpha}$ have a linear correction in $D$
(see eq. (\ref{transformation})) that we have neglected so far. The
intensity at $\mathbf{q}=(\pi,0)$ (that was zero before) acquires a
term proportional to $(D/J)^2$: $S^0((\pi,0)) = \frac{D^2}{8J^2}
N_{\mathbf{q}}$ where $N_{\mathbf{q}}=\cos^2 \mathbf{q} \cdot
\mathbf{\delta} + \cos^2 \mathbf{q} \cdot
\mathbf{\delta}^{\prime}$. So in principle it is possible to extract
$D$ from measuring the ratio of intensities at $\mathbf{q}=(\pi,0)$,
$\frac{S^0((\pi,0))}{S^{\pm}((\pi,0))} = \frac{D^2}{8J^2} \tan^2
\frac{\pi}{6 \sqrt{2}} $.  This ratio is very small $\sim 0.02
(D/J)^2$, and it therefore seems difficult to extract this coupling
independently.  Second, we consider the effect on the intensities of
the non-staggered components
$D_{\parallel,ns}^{\prime}+J^{\prime}D/2J$, that we had eliminated by
the second rotation. Since the rotation is performed using a constant
angle when going from dimer to dimer, the spin operators
$S_{\mathbf{q}}^{\alpha}$ contains operators that are shifted in
$\mathbf{q}$ space, $S_{\mathbf{q} \pm \mathbf{k}_0}^{\alpha}$, where
$\mathbf{k}_0$ is determined by the coupling
$D_{\parallel,ns}^{\prime}+J^{\prime}D/2J$. A consequence is that
``ghosts'' of the original magnon dispersion (as given in
Fig.~\ref{DM}) shifted by $\pm \mathbf{k}_0$ must appear in the
spectrum. Of course because the shift is small, of order $\lambda$,
and the dispersion is of the same order, the energies are changed by
$\lambda^2$. The effect will be difficult to detect by neutron
scattering using constant wave-vector scans, but could appear in
constant energy scans.  This is reminiscent of magnon dispersions in
well ordered systems, such as Ba$_2$CuGe$_2$O$_7$, where the \DM
interaction and the helicoidal magnetic structure give rise to three
branches in the excitation spectrum.\cite{Zheludev}

\textit{Singlet states.} In Figs.~\ref{DM} and \ref{D}, higher and
lower \textit{singlet} states are also given. Since the rotational
symmetry is broken, denoting a state as a ``singlet''is to be
understood as its property for vanishing anisotropies. Singlets are
not directly visible by ESR, because the external field is swept at
constant wavelengths and only field-dependent states are
detected. Nonetheless the bending of the triplet energy levels tend to
signal a level anti-crossing with a singlet: experimentally such
bendings were observed and thus singlet energies indirectly
measured.\cite{NojiriESR1} The measured anti-crossings are at 2.67 meV
(compared to the lowest triplet state at 2.81 mev) and 3.56 meV
(compared to the higher triplet at 3.16 meV).\cite{NojiriESR1} In
Fig.~\ref{DM} the lowest singlet has an energy which is 95\% of the
triplet gap, in good agreement with the experimental figure. However
the second singlet seems to be higher in energy compared to that of
Fig.~\ref{DM}. We know that there are other singlet states at higher
energies, and that observed by ESR may be one of them.

\textit{Sign of the coupling}.
We remark that although the sign of $D_{\perp}^{\prime}$ is of no
consequence on the energy spectrum, it interchanges the ``handedness''
of the wave functions of the upper and lower states. 
It therefore has measurable consequences when polarized light interacts with the
triplets.\cite{CepasOptical} It is interesting to note that the sign
of $D_{\perp}^{\prime}$ extracted by analyzing\cite{CepasOptical}
far-infrared experiments\cite{Room1,Room2} is opposite to that
suggested to explain avoided crossing at higher
fields.\cite{Miyaharaintra,Kodama} The apparent contradiction may be
resolved because the avoided crossing is not determined solely by $D$,
as suggested earlier\cite{Miyaharaintra,Kodama,MiyaharaQSS04}. but
rather by a combination of $D_{\parallel,s}^{\prime}+J^{\prime}D/2J$,
$D_{\parallel,ns}^{\prime}+J^{\prime}D/2J$ and the staggered
fields. Quantitative understanding  of the avoided crossing would
need determination of each individual coupling.

\section{Conclusions}

We have considered a model for SrCu$_2$(BO$_3$)$_2$ with all the \DM
 interactions compatible with the crystal structure, \textit{i.e.} a
 model with all anisotropic couplings linear in the spin-orbit
 coupling. We have constructed a simpler anisotropic Hamiltonian
 (\ref{hammapped1}) with a smaller number of couplings, by appropriate
 mappings, which  has nonetheless exactly the same spectrum up to second-order
 interactions.  The transformation allowed us to separate what we term
 the ``reducible'' components of \DM vectors (the non-staggered
 components $D_{\parallel,ns}^{\prime}$ and part of the intra-dimer
 interaction $D$) from the ``irreducible'' (the staggered
 $D_{\perp}^{\prime}$, $D_{\parallel,s}^{\prime}$ with a contribution from the
 intra-dimer interaction $D$). By definition the latter have effects
 on the spin excitation spectrum that are linear in the strength of
 the coupling, whereas the former have second-order effects. The
 linear effect arises, precisely, because we cannot eliminate these
 components. Had we been able to eliminate all of them, we should 
 have taken into account anisotropic symmetric {\it exchanges}, on an equal
 footing.\cite{rotation} In that case, depending on the underlying
 superexchange processes, one may  recover a fully rotationally
 invariant spectrum (if single-orbital processes are
 dominant),\cite{rotation} or not (when multi-orbital processes
 exist).\cite{Maekawa} In any case, the effects would have been at
 most $\lambda^2$, and very difficult to extract experimentally
 because this would be of order of a few $\mu$eV. In contrast,
 SrCu$_2$(BO$_3$)$_2$ represents a real situation where the bond
 frustration leads to an energy scale of order $\lambda$ -hundreds of
 $\mu$eV here- in the spin excitation spectrum.

We have explained quantitatively the dispersion of the lowest triplet
states (Fig.~\ref{DMexp}) which is dominated by \DM interactions
(compare Fig.~\ref{noDM} and \ref{DM}). First-order perturbation theory was not sufficient to
 account quantitatively for the dispersion shape (see Fig.~\ref{noJ})
but allows us to capture most of the symmetries except for the small
additional splittings due to the presence of two dimers per
unit-cell. Measurements of the dispersion under high-pressure\cite{KageyamaUnpub} should
in fact be carefully interpreted: the evolution of the bandwidth
with pressure should primarily reflect the change in the \DM couplings
and not the change in the ratio $J^{\prime}/J$ that measure the
proximity to the critical point.  The spin gap, where the effect of
\DM is secondary, may serve as a more sensitive probe for the
proximity to the critical point.

The model (\ref{hammapped1}) whose parameters have been quantitatively
determined, may serve as a good starting point to understand other
properties of SrCu$_2$(BO$_3$)$_2$ at zero field, such as higher
energy bound states.  These states, as observed in particular by
ESR,\cite{NojiriESR1} Raman scattering,\cite{Lemmens,Gozar}
far-infrared spectroscopy,\cite{Room2} and neutron scattering
\cite{Kakurai0,NewKakurai} have been analyzed so far within the
framework of the Shastry-Sutherland
model,\cite{Fukumoto,Totsuka,KnetterUhrig} but 
 should be affected by \DM
interactions. This may explain the splittings
observed,\cite{NewKakurai} and resolve the issue of ``localization''
versus ``delocalization'' of these excitations. The large dispersion
seen first in ref. [\onlinecite{Kakurai0}] and supported by
theoretical calculations on the Shastry-Sutherland model was revealed
latter to be composed of several more localized
excitations.\cite{NewKakurai} This remains to be fully understood
and model (\ref{hammapped1}) may help.

We have focused on zero-field results: finite field effects are
particularly interesting but involve other unknown couplings.  If the
mappings of section~\ref{mapping} (transformations
(\ref{transformation}) and (\ref{transformation2})) are made with an
external magnetic-field, both an effective staggered
field,\cite{Oshikawa} and rotating fields appear. As a real staggered
field is already present according to the local symmetry of the copper
ions, the net effective fields cannot be known precisely. The level
anti-crossing that was found when the spin gap is about to close for a
magnetic-field along the $c$ direction, in particular, cannot be
solely due to the intra-dimer interaction, as previously
claimed.\cite{Miyaharaintra} It has to be seen as a consequence of all
these effects: intra and inter-dimer components, and staggered fields.

Interesting developments for doped samples of SrCu$_2$(BO$_3$)$_2$ are
underway.\cite{Liu,Haravifard} The doped compounds, by breaking
translation invariance, may also help to disentangle the different
components of the \DM vectors, by studying the induced magnetization
of the magnetic ions in the vicinity of the dopant, by NMR for
instance.

\acknowledgments

We are indebted to K. Kakurai for providing with us the experimental
data, and for constant encouragement along these lines.  We would also
like to acknowledge the role of J.-P. Boucher in stimulating our
interest in this subject.  O.C. and T.Z. would like to thank HKUST for
hospitality on several occasions; and O.C. the I.L.L. for hospitality
for part of the time. T.Z. and P.W.L. would like to thank the PROCORE
program for support which lead to this collaboration. Y.F.C. and
P.W.L. are supported by the Hong Kong RGC Grant No. HKUST6075/02P.

\end{document}